\documentclass{svproc}

\usepackage[T1,hyphens]{url}
\usepackage{graphicx}
\usepackage{subfigure}
\usepackage{multirow}
\usepackage{amsmath} 
\usepackage{stmaryrd}
\usepackage{float}
\usepackage{rotating}
\usepackage{multirow}
\usepackage[titletoc, title]{appendix}
\usepackage{adjustbox}
\usepackage{bbm}
\usepackage{color}
\usepackage[table]{xcolor}
\usepackage{lscape}
\usepackage{tablefootnote}
\usepackage{subfig} 
\usepackage{caption}
\usepackage{morefloats}

\usepackage{tablefootnote}

\usepackage[utf8]{inputenc}

\usepackage{array} 
\usepackage{tabularx} 
\newcolumntype{Y}{>{\normalsize\arraybackslash}X} 

\usepackage[misc,geometry]{ifsym}

\usepackage{comment}

\usepackage[square,numbers]{natbib} 
\makeatletter
\renewcommand\bibsection%
{
 \section*{\refname
 \@mkboth{\MakeUppercase{\refname}}{\MakeUppercase{\refname}}}
}
\makeatother

\usepackage{stfloats}

\usepackage{todonotes}

\usepackage{multicol,booktabs}

\begin{document}

\title{Epidemic Information Extraction for Event-Based Surveillance using Large Language Models} 

\titlerunning{Epidemic Information Extraction using LLMs}

\author{Sergio Consoli\inst{1}\Letter \and  Peter Markov\inst{1} \and Nikolaos I. Stilianakis\inst{1} \and Lorenzo Bertolini\inst{1} \and  Antonio Puertas Gallardo\inst{1} \and Mario Ceresa\inst{1}}

%
\authorrunning{Consoli et al.} 
%
\institute{European Commission, Joint Research Centre (JRC), Ispra, Italy. 
\email{ [name.surname]@ec.europa.eu}
}

\maketitle 

\begin{abstract}
This paper presents a novel approach to epidemic surveillance, leveraging the power of Artificial Intelligence and Large Language Models (LLMs) for effective interpretation of unstructured big data sources, like the popular ProMED and WHO Disease Outbreak News. We explore several LLMs, evaluating their capabilities in extracting valuable epidemic information. We further enhance the capabilities of the LLMs using in-context learning, and test the performance of an ensemble model incorporating multiple open-source LLMs. The findings indicate that LLMs can significantly enhance the accuracy and timeliness of epidemic modelling and forecasting, offering a promising tool for managing future pandemic events.
\keywords{Health Informatics; Epidemiology; Event-based Surveillance; Natural Language Processing; Large Language Models. 
} 
\end{abstract}
%

\section{Introduction and background} \label{introduction} 

Epidemic modelling 
is a challenging task as emphasized by numerous studies \citep{vespignani2011}. 
The use of novel unstructured data sets 
has been increasingly encouraged, as they can significantly contribute to improving early warning systems \citep{salathe2012}. 
%
The recent Covid-19 pandemic has underscored the importance of these novel approaches \citep{McDonald2021}. The pandemic revealed the long delays in the release of official statistics, emphasizing the need to track pandemic information in alternative ways and at a higher frequency. 

These opportunities and challenges in infectious disease epidemiology 
are inspiring some of the research activities at the Digital Health unit of the Joint Research Centre (JRC)\footnote{\url{https://ec.europa.eu/info/departments/joint-research-centre_en}} of the European Commission (EC). Ongoing work, described in this paper, aims at tracking pandemic outbreaks in the European Union (EU) using 
data sets which are considered unconventional in classic epidemiological modeling. The project aims at exploring novel (big) data sources to provide a better response to epidemics. 
It 
links to other ongoing relevant initiatives on the subject\footnote{WHO EIOS: Epidemic Intelligence from Open Sources, \url{https://www.who.int/initiatives/eios}}.
%
%
An important source of such information is represented by moderated news and reports, 
since they discuss important events and experts opinions, and 
can substantively inform policy-making decisions \citep{Leuba2020}. However, translating this new source of data into valuable information is challenging, given that the data derived from these sources are often unstructured and large. This data also exhibits non-linear relationships across variables, which adds to the complexity of its interpretation. Despite these challenges, the potential benefits of utilizing these unconventional data sources are considerable. By effectively translating this data into actionable insights, we can significantly improve our understanding of epidemics and, consequently, our response to them. 

In this paper, we aim to leverage the vast capabilities of Artificial Intelligence (AI) \citep{Brownstein20231597}, specifically the power of Generative Pre-trained Large Language Models (LLMs) \citep{Vaswani20175999,Brown2020}, for the the exploration and effective interpretation of such innovative big data sources, with the ultimate goal of improving epidemic response. 
LLMs are a type of generative AI models that utilizes deep-learning (DL) algorithms (involving billions of parameters) to calculate the likelihoods of word sequences. These probabilities are determined based on substantial text corpora that the model has previously learned from. Significant advancements to LLMs have been made with the advent of the Transformers architecture \citep{Vaswani20175999}. Transformers 
designed to handle sequential data by using a mechanism called attention, which allows the model to weigh and prioritize different parts of the input data \citep{sutskever2014sequence}. Unlike traditional sequential models such as Recurrent Neural Networks (RNNs), Transformers process all input data concurrently, which allows for more efficient computation and the ability to handle longer sequences. 

Considering the rapid advancements in LLMs, this study aims to evaluate the most significant and recent open-source and commercial models for the specific task of epidemic information extraction. The use of the information extracted from large-scale, unstructured data sets in epidemic infectious disease, coupled with the integration to surveillance systems, can significantly enhance the accuracy and timeliness of epidemic outbreaks modelling and forecasting. 


\section{Data} \label{data} 

In this section, we describe the unstructured data sets used in the study. 
Although we have focused on these two data sets, the approach is completely generalizable to any textual source concerning infectious disease epidemics. 

\subsection{ProMED}  \label{promed} 

ProMED\footnote{\url{https://promedmail.org/}}, the Program for Monitoring Emerging Diseases, is an initiative by the International Society for Infectious Diseases (ISID). As the largest publicly accessible system for infectious disease outbreak reporting, it serves as a critical resource for healthcare professionals and the public alike. Its platform offers a wealth of daily posts detailing the latest developments in the field of infectious diseases.
Launched in 1994, ProMED has been at the forefront of outbreak reporting, being the first to report on numerous disease outbreaks, including SARS, Chikungunya, Ebola, Zika, MERS, and most recently, COVID-19 \citep{madoff2005internet}. Its users comprise international public health leaders, government officials, physicians, veterinarians, researchers, companies, journalists, and the general public worldwide. 
%
Reports are produced and commentary provided by a multidisciplinary global team of subject matter experts 
in a variety of fields, including virology, parasitology, epidemiology, and entomology. Since its inception, ProMED has generated so far more than 66,000 
mail posts, from 1994-08-19 to date.

\subsection{WHO Disease Outbreaks News}  \label{dons} 

The World Health Organization (WHO)'s Disease Outbreak News (DONs)\footnote{\url{https://www.who.int/emergencies/disease-outbreak-news/}} serve as a vital source of information on confirmed acute public health events or potential events of concern. The platform 
has been providing crucial updates since January 1996. 
The essence of DONs lies in its commitment to sharing news about confirmed or potential public health events. This includes incidents of unknown cause that carry significant or potential international health concerns and could affect international travel or trade. It also encompasses diseases of known cause which have shown the capacity to cause a serious public health impact and spread internationally. 
Since its inception, DONs has published over 3,000 pieces of curated epidemic news, from 1996-01-22 to date. 

\section{Methods}  \label{Models} 

In this section, we describe the epidemic information extraction models considered in our study. 
These consist in the most popular and recent LLMs, spanning from open-source to commercial ones, along with a significant semantic method from the literature for epidemic information extraction, namely EpiTator.
%
Please note that the LLMs employed in this study have been used through the GPT@JRC initiative of the Joint Research Centre (JRC) of the European Commission, which enables JRC staff to explore the potential uses of AI pre-trained LLMs. The initiative, which is part of a broader study on the new technology's applications within the European Commission, is a central hub offering secure access to various AI models. 
GPT@JRC is hosted at the JRC datacentre and supports both open-source AI models, deployed on-premises at the JRC Big Data Analytics Platform\footnote{\url{https://jeodpp.jrc.ec.europa.eu/bdap/}} 
and commercial OpenAI's GPT models running in the European Cloud under a Commission contract with 
an opt-out clause on prompt analysis by third parties.

\subsection{EpiTator}  \label{EpiTator} 

EpiTator\footnote{\url{https://github.com/ecohealthalliance/EpiTator}} is a comprehensive open-source epidemiological annotation tool, specifically designed for the extraction of epidemiological information from texts\citep{Abbood2020}. This tool is essentially a Python script that leverages the powerful Natural Language Processing (NLP) spaCy library\footnote{\url{https://spacy.io/}} to extract critical named-entities that are of particular interest in epidemiology applications. These entities include diseases, locations, dates, and counts. 
One of the remarkable features of EpiTator is the \emph{Resolved Keyword Annotator}. It utilizes an SQLite database of entities, which ensures that multiple synonyms for an entity are resolved to a single id, thereby enhancing the accuracy and consistency of the extracted data. EpiTator also imports information about infectious diseases and animal species from several reliable sources, including Disease Ontology, Wikidata, and the Integrated Taxonomic Information System (ITIS). 
The \emph{Count Annotator} feature of EpiTator extracts and parses count values, identifying attributes such as whether the count refers to cases or deaths, or if the value is approximate. This provides a more nuanced understanding of the data. Instead, the \emph{Date Annotator} feature identifies and parses dates and date ranges. 
EpiTator employs a key entity filtering process that condenses the output to a single entity per class that best describes the corresponding event. Despite EpiTator’s capability to return all entities of an entity class (e.g., disease) found in a text, this filtering process is necessary to distill the most pertinent information. It achieves this using the most-frequent approach filtering, which identifies the key entity from all the entities by selecting the most frequently mentioned one per class. 

\subsection{Pythia-12b}  \label{Pythia-12b} 

The Pythia-12b model is a state-of-the-art transformer-based LLM that follows the architecture of the highly popular GPT3 model. This open-source model is owned by EleutherAI\footnote{\url{https://huggingface.co/EleutherAI/pythia-12b}}, a collective of AI specialists aiming at advancing AI in an open and collaborative manner. Pythia-12b has been primarily designed to facilitate research on the behaviour, functionality, and limitations of LLMs. 
The model operates within a context length of 4,096 tokens and consists of a whopping 12 billion parameters. 

\subsection{Mpt-30b-chat}  \label{Mpt-30b-chat} 

The Mpt-30b-chat model is a state-of-the-art, open-source LLM owned by MosaicML\footnote{\url{https://huggingface.co/mosaicml/mpt-30b-chat}}. 
One of the key features 
is its specialization in chat and dialogue generation. It has been fine-tuned to handle conversational dynamics, making it an excellent choice for creating interactive AI applications. 
It was fine-tuned on several datasets including ShareGPT-Vicuna, Camel-AI, GPTeacher, Guanaco, Baize, and some more generated datasets. 
In terms of technical specifics, the Mpt-30b-chat model boasts a context length of 8,192 tokens, 
and is built with an impressive 30 billion parameters. This vast number of parameters allows the model to capture and learn from a wide variety of linguistic nuances, thereby greatly enhancing its conversational capabilities and predictive performance. 

\subsection{Llama-2-70b-chat}  \label{Llama-2-70b-chat} 

The Llama-2-70b-chat model \citep{touvron2023llama} is an advanced open-source LLM owned by Meta\footnote{\url{https://ai.meta.com/llama/}}. It is one of the most powerful open LLMs currently available. 
The base model, Llama 2, was pretrained on a diverse range of publicly available online data sources. 
The fine-tuned version of this model, known as Llama-2-70b-chat, further enhances this understanding by leveraging publicly available instruction datasets and over 1 million human annotations. This extensive fine-tuning process ensures a high level of precision and adaptability in its responses. 
In terms of its technical specifications, the Llama-2-70b-chat model operates with a context length of 4,096 tokens. 
Moreover, the model is built with an impressive 70 billion parameters. 

\subsection{Mistral-7b-openorca}  \label{Mistral-7b-open} 

The Mistral-7b-openorca model\footnote{\url{https://huggingface.co/Open-Orca/Mistral-7B-OpenOrca}} is an open-source model, owned and fine-tuned by OpenOrca using its datasets\footnote{\url{https://huggingface.co/datasets/Open-Orca/OpenOrca}} on top of the Mistral LLM \citep{mukherjee2023orca}. Despite its smaller size with 7 billion parameters, the model is fast at inference and delivers robust performance across a wide range of tasks. 
With a context length of 4,096 tokens, the Mistral-7b-openorca is one of the best performing open-source models available for models under 30 billion parameters. This makes it a strong candidate for many API use-cases due to its combination of size, speed, and performance.

\subsection{Zephyr-7b-alpha}  \label{Zephyr-7b-alpha} 
The Zephyr-7b-alpha model is an open-source model owned by HuggingFace\footnote{\url{https://huggingface.co/HuggingFaceH4/zephyr-7b-alpha}}. This model has been fine-tuned by HuggingFace using a combination of publicly available and synthetic datasets on top of the Mistral LLM. 
Despite its small size with only 7 billion parameters, it often outperforms the larger Llama 2 13B model, demonstrating performance comparable to several models in the 20-30B range. 

\subsubsection{Gpt-3.5-turbo-16k}  \label{Gpt-3.5-turbo-16k} 

The Gpt-35-turbo-16k model is a robust commercial model owned by OpenAI. This model is an advanced version of OpenAI's GPT-35-turbo, colloquially known as ChatGPT, but with four times the context. 
The versatility of this model is apparent in its ability to perform a wide range of tasks. 
With its context length of 16,384 tokens, the Gpt-35-turbo-16k model offers an expansive reach for various applications. However, being a commercial model, its extensive usage comes with a considerable cost and some usage 
constraints.

\subsection{Gpt-4-32k}  \label{Gpt-4-32k} 

The Gpt-4-32k model, another commercial offering from OpenAI, is a powerful tool which sets itself apart in terms of its problem-solving capabilities. It can solve difficult problems with a greater level of accuracy than any of OpenAI's previous models, leading to its reputation as the best LLM available for any task. While it may come at a higher cost, the Gpt-4-32k model's superior performance and its ability to manage approximately 80 pages of text make it one of the best LLM option available. 
Its context length of 32,768 tokens provides an even greater scope for processing and managing large amounts of data, making it a perfect option to handle large texts.

\subsection{Gpt-4-FewShots}  \label{Gpt-4-FewShots} 

The Gpt-4-FewShots is a variant to the Gpt-4-32k model obtained by passing three examples (Three-Shots) of epidemic information extraction in the context to the Gpt-4-32k prompt, such that to try to specialize the model 
to handle the specific epidemic information extraction task. 
In-context learning (ICL) \citep{Brown2020} refers to a learning approach where an AI model learns from the its context. LLMs have already shown an ability for ICL as they scale in terms of model and corpus sizes \citep{dong2023survey}. 
The fundamental principle of in-context learning is learning through analogy. Initially, ICL needs a handful of examples to create a demonstration context, usually framed in natural language. Following this, ICL combines a query question with a demonstration context to create a prompt. This prompt is then input into the language model for prediction.
%
%
We have leveraged the ICL approach on the Gpt-4-32k model\footnote{Please note that ICL was only applied to Gpt-4-32k due its large context capability, which was not amenable in the other studied LLMs which are characterised by a way more lower context.}, leading to Gpt-4-FewShots, to allow for a more efficient and accurate way to extract and process information on epidemics. 

\subsection{Open-Ensemble}  \label{Open-Ensemble} 

The Open-Ensemble model is an AI approach that leverages the power of multiple high-performing open-source models to generate outputs. It uses a majority voting system to determine the best result from the contributed models. 
In contrast to ordinary machine learning approaches which try to learn one hypothesis from training data, ensemble methods try to construct a set of hypotheses and combine them to use \citep{Sagi2018}.

The models included in this ensemble are Llama-2-70b-chat, Mistral-7b-openorca, and Zephyr-7b-alpha, that, as it will be shown next, resulted to be the best performing open-source LLMs for epidemic IE. Each of these models bring unique strengths and capabilities to the ensemble, enhancing its overall performance. 
This Open-Ensemble model, with its majority voting mechanism, ensures that the most agreed-upon output from these three models is chosen, thereby increasing the likelihood of accuracy and reliability in its predictions. 
We have used this approach to test whether the obtained ensemble model improves over the single open-source LLMs, and is able to compete with the performance of the more advanced commercial Gpt models.

\section{Computational experiments} \label{expe} 

In this section, we delve into the application of LLMs for the purpose of epidemic information extraction. 
%
The extracted entities include the name of the disease, the country where the cases were reported, the confirmed case count, and the date of the case count. 
We compare the performance of the different LLMs for infectious disease epidemic information extraction, along with EpiTator. 
The comparison has been made over a subset of the Incident Database (IDB), a repository developed 
for the purpose of event-based surveillance\footnote{\url{https://github.com/aauss/EventEpi}}. The IDB is structured to house a comprehensive collection of data related to epidemic events, making it an important resource for researchers and healthcare professionals.
The database includes a sample of ProMED and WHO DONs that have been meticulously annotated by domain experts. 
For the purpose of comparing the performance of the tested epidemic information extraction models, a gold standard subset of the IDB comprising 171 carefully selected samples has been selected. This subset provides a benchmark against which the performance of the models have been assessed.

We have evaluated the information extraction task as a binary classification problem where the negative class means that no information (``None'') is associated to the IDB data sample, while the the positive class indicates that the IDB is labelled with a specific epidemic information. The models have been tested on their ability to recognize such positive and negative classes by considering the following 
%
%
widely adopted Precision, Recall, and $F_1$ score metrics~\citep{DBLP:books/sp/CRP2019}: 
%
$Precision = \frac{TP}{TP+FP}$, $Recall = \frac{TP}{TP+FN}$, $F_1 = \frac{2}{\frac{1}{Precision}+\frac{1}{Recall}}$, where $TP$, $FP$, $FN$, and $TN$ are, respectively, the number of True Positives, the number of False Positives, the number of False Negatives, and the number of True Negatives~\citep{DBLP:books/sp/CRP2019}.
In words, the Precision is the accuracy for the positive class predictions. The Recall (Sensitivity) is the ratio of the positive class instances that are correctly detected as such by the classifier. The $F_1$ score is the harmonic mean of precision and recall: whereas the regular mean treats all values equally, the harmonic one gives much weight to low values, and for this reason is a metric preferred to the classical accuracy in an imbalanced problem setting. 


\begin{figure}[!t]
    \begin{minipage}[c]{0.5\textwidth}
    \hspace*{-1.2cm}
\begin{tabularx}{\linewidth}{lp{0.1cm}p{1.18cm}p{1.18cm}p{1.18cm}}
\toprule 
							& &	\textsf{Prec}	&	\textsf{Rec}	&	\textsf{F1}	\\
\hline	
\textsf{EpiTator}			& &	0.692	&	0.719	&	0.705  \\
\textsf{Pythia-12b}			& &	0.712	&	0.773	&	0.742  \\
\textsf{Mpt-30b-chat}		& &	0.74	&	0.891	&	0.809  \\
\textsf{Llama-2-70b-chat}	& &	0.757	&	0.898	&	0.821  \\
\textsf{Mistral-7b-open}	& &	0.755	&	0.891	&	0.817  \\
\textsf{Zephyr-7b-alpha}	& &	0.752	&	0.875	&	0.809  \\
\textsf{Gpt-35-turbo-16k}	& &	0.75	&	0.961	&	0.842  \\
\textsf{Gpt-4-32k}			& &	0.756	&	0.945	&	0.84   \\
\hline \hline
\textsf{Gpt-4-FewShots}		& &	0.77	&	0.914	&	0.836  \\
\textsf{Open-Ensemble}		& &	0.756	&	0.921	&	0.83   \\
\bottomrule
\end{tabularx} 
    \end{minipage}
    \hfill
    \begin{minipage}[c]{0.7\textwidth}
    \hspace*{-1.1cm}
        \includegraphics[width=\linewidth]{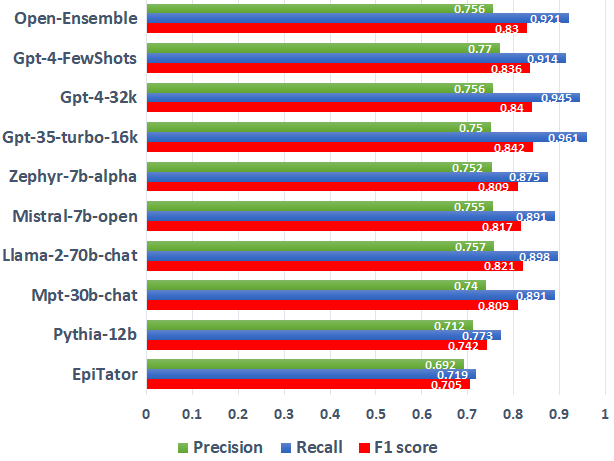}
    \end{minipage}
    \caption{Comparison of the models for the extraction of the pandemic name.}
    \label{pandemic}
\end{figure} 

\begin{figure}[!ht]
    \begin{minipage}[c]{0.5\textwidth}
    \hspace*{-1.2cm}
\begin{tabularx}{\linewidth}{lp{0.1cm}p{1.18cm}p{1.18cm}p{1.18cm}}
\toprule 
							& &	\textsf{Prec}	&	\textsf{Rec}	&	\textsf{F1}	\\
\hline	
\textsf{EpiTator}			&	&	0.978	&	0.819	&	0.892	\\
\textsf{Pythia-12b}			&	&	0.962	&	0.452	&	0.615	\\
\textsf{Mpt-30b-chat}		&	&	0.979	&	0.855	&	0.913	\\
\textsf{Llama-2-70b-chat}	&	&	0.98	&	0.898	&	0.937	\\
\textsf{Mistral-7b-open}	&	&	0.986	&	0.861	&	0.92	\\
\textsf{Zephyr-7b-alpha}	&	&	0.981	&	0.91	&	0.944	\\
\textsf{Gpt-35-turbo-16k}	&	&	0.981	&	0.91	&	0.944	\\
\textsf{Gpt-4-32k}			&	&	0.981	&	0.928	&	0.954	\\
\hline \hline
\textsf{Gpt-4-FewShots}		&	&	0.981	&	0.928	&	0.954	\\
\textsf{Open-Ensemble}		&	&	0.981	&	0.91	&	0.944	\\	
\bottomrule
\end{tabularx}
    \end{minipage}
    \hfill
    \begin{minipage}[c]{0.7\textwidth}
    \hspace*{-1.1cm}
        \includegraphics[width=\linewidth]{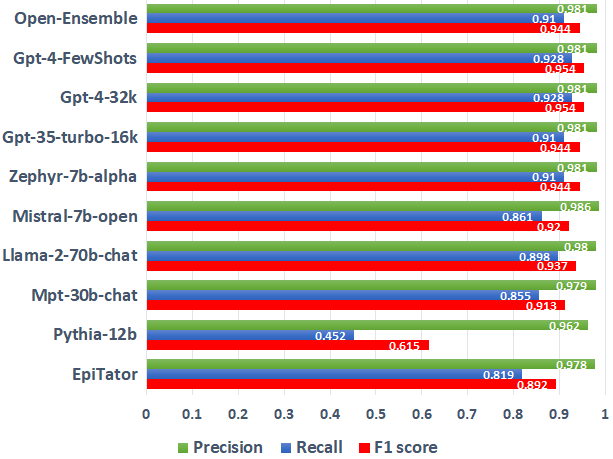}
    \end{minipage}
    \caption{Comparison of the models for the extraction of the country name.} \label{country}
\end{figure}

\begin{figure}[!ht]
    \begin{minipage}[c]{0.5\textwidth}
    \hspace*{-1.2cm}
\begin{tabularx}{\linewidth}{lp{0.1cm}p{1.18cm}p{1.18cm}p{1.18cm}}
\toprule 
& &	\textsf{Prec}	&	\textsf{Rec}	&	\textsf{F1}	\\
\hline							
\textsf{EpiTator}	& &	0.892	&	0.576	&	0.7	\\
\textsf{Pythia-12b}	& &	0.735	&	0.158	&	0.26	\\
\textsf{Mpt-30b-chat}	& &	0.814	&	0.304	&	0.442	\\
\textsf{Llama-2-70b-chat}	& &	0.916	&	0.759	&	0.83	\\
\textsf{Mistral-7b-open}	& &	0.908	&	0.69	&	0.784	\\
\textsf{Zephyr-7b-alpha}	& &	0.92	&	0.804	&	0.858	\\
\textsf{Gpt-35-turbo-16k}	& &	0.902	&	0.639	&	0.748	\\
\textsf{Gpt-4-32k}	& &	0.913	&	0.734	&	0.814	\\
\hline \hline							
\textsf{Gpt-4-FewShots}	& &	0.904	&	0.658	&	0.762	\\
\textsf{Open-Ensemble}	& &	0.92	&	0.804	&	0.858	\\
\bottomrule
\end{tabularx}
    \end{minipage}
    \hfill
    \begin{minipage}[c]{0.7\textwidth}
    \hspace*{-1.1cm}
        \includegraphics[width=\linewidth]{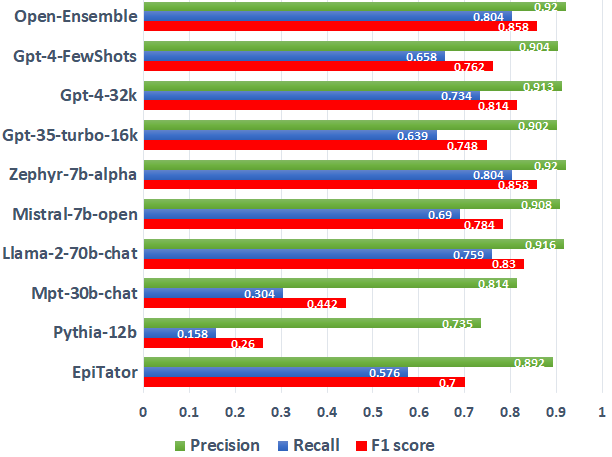}
    \end{minipage}
    \caption{Comparison of the models for the extraction of the pandemic date.} \label{date}
\end{figure}

\begin{figure}[!ht]
    \begin{minipage}[c]{0.5\textwidth}
    \hspace*{-1.2cm}
\begin{tabularx}{\linewidth}{lp{0.1cm}p{1.18cm}p{1.18cm}p{1.18cm}}
\toprule 
& &	\textsf{Prec}	&	\textsf{Rec}	&	\textsf{F1}	\\
\hline							
\textsf{EpiTator}			&	&	0.387	&	0.321	&	0.351	\\
\textsf{Pythia-12b}			&	&	0.365	&	0.277	&	0.315	\\
\textsf{Mpt-30b-chat}		&	&	0.619	&	0.464	&	0.531	\\
\textsf{Llama-2-70b-chat}	&	&	0.561	&	0.536	&	0.548	\\
\textsf{Mistral-7b-open}	&	&	0.67	&	0.527	&	0.59	\\
\textsf{Zephyr-7b-alpha}	&	&	0.615	&	0.571	&	0.593	\\
\textsf{Gpt-35-turbo-16k}	&	&	0.699	&	0.455	&	0.551	\\
\textsf{Gpt-4-32k}			&	&	0.673	&	0.589	&	0.629	\\
\hline \hline
\textsf{Gpt-4-FewShots}		&	&	0.733	&	0.589	&	0.653	\\
\textsf{Open-Ensemble}		&	&	0.625	&	0.581	&	0.601	\\
\bottomrule
\end{tabularx}
    \end{minipage}
    \hfill
    \begin{minipage}[c]{0.7\textwidth}
    \hspace*{-1.1cm}
        \includegraphics[width=\linewidth]{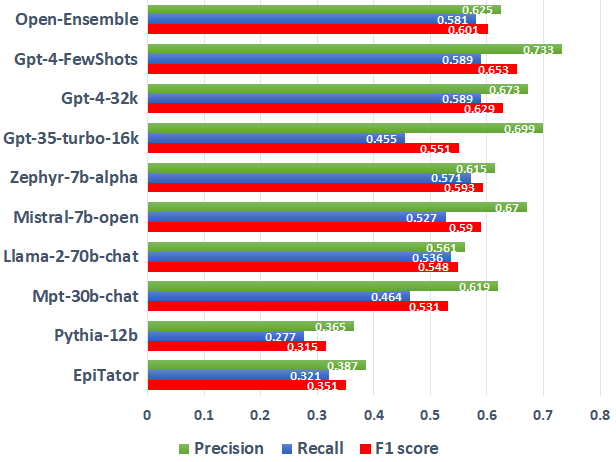}
    \end{minipage}
    \caption{Comparison of the models for the extraction of the number of cases engender by the pandemic.} \label{case}
\end{figure}

Our computational results are reported in Figure \ref{pandemic}, Figure \ref{country}, Figure \ref{date} and Figure \ref{case}, which report the comparison of the different algorithms for the extraction, respectively, of the pandemic name, the country where this virus outbreak occurred, the related date, and the number of cases associated to the specific outbreak. In particular, each illustration report the considered metric scores obtained by each method (on the left), also with the corresponding graphical performance illustration, for a better understanding (on the right). 
Focusing on the performance of the LLMs, two models, namely Pythia-12b and Mpt-30b-chat, have shown underwhelming performance, falling below the standard NLP-based method, EpiTator. However, this was not the case for all LLMs. The majority of them have demonstrated a remarkable ability to outperform EpiTator by a significant margin. Among the LLMs in the evaluation, Gpt-4-32k stood out by achieving the best overall performance, as expected. Another very well performer was Gpt-35-turbo-16k, which also managed to achieve a significant performance level, as shown in the tables of the results. Furthermore, experimenting with in-context learning via the 3-shots method (Gpt-4-32k-FewShots), we have managed to improve the performance of Gpt-4, although this was not always the case. 

Although the superior performance of OpenAI gpt models relative to the other tested methods, their use 
comes with its own set of challenges. These models are costly to implement, and some of their usage restrictions 
make them unsuitable in practice for a extensive deployment on a large dataset, like the full ProMED and DONs. On the brighter side, some open-source models performed very well, albeit slightly below the level of the OpenAI GPTs. These include Llama-2-70b-chat, Mistral-7b-openorca, and Zephyr-7b-alpha. It's important to note that Llama-2-70b-chat was found to be slower in terms of computational times compared to the rest, given with large number of parameters (70 billions) of its underlying model. 
Finally, the adopted ensemble approach (OpenLLMs-Ensemble) on these three open-source LLMs produces 
an overall robust and satisfactory performance on all the four epidemic information extraction tasks, resulting to be comparable to the gpt models.
Therefore, OpenLLMs-Ensemble allows for a full deployment on the entire ProMED and DONs data, spanning approximately 70,000 documents, thereby providing a comprehensive and efficient solution forinfectious disease epidemic information extraction.

\section{Conclusions} \label{conclusions} 

This study has demonstrated the significant potential of LLMs in epidemic surveillance, particularly in the extraction of valuable epidemic information from unstructured big data sources. Several LLMs were evaluated and their capabilities in extracting epidemic information were assessed. The findings of this study suggest that LLMs, particularly when used in an ensemble, can significantly enhance the accuracy and timeliness of epidemic modelling and forecasting.
%
%
This application of LLMs not only streamlines the data gathering process but also has the potential to improve 
early warning systems and epidemic response strategies. 
These models provide a promising tool for managing future pandemic events, 
reducing their impact on society and economies, and, finally, saving lives.


\section*{Acknowledgements}
The views expressed are purely those of the authors and do not, in any circumstance, be regarded as stating an official position of the European Commission. 
We would like to acknowledge the GPT@JRC initiative for providing access to the LLMs used in this study. 
We extend our gratitude also to the JRC Big Data Analytics Platform for 
providing secure access to various open-source AI models. 
Finally, we would like to thank the colleagues contributing to the WHO EIOS initiative for the helpful suggestions during the development of this work.


%
%
\bibliographystyle{abbrvnat}

\bibliography{bibliofile}

\begin{thebibliography}{15}
\providecommand{\natexlab}[1]{#1}
\providecommand{\url}[1]{\texttt{#1}}
\expandafter\ifx\csname urlstyle\endcsname\relax
  \providecommand{\doi}[1]{doi: #1}\else
  \providecommand{\doi}{doi: \begingroup \urlstyle{rm}\Url}\fi

\bibitem[Abbood et~al.(2020)Abbood, Ullrich, Busche, and Ghozzi]{Abbood2020}
A.~Abbood, A.~Ullrich, R.~Busche, and S.~Ghozzi.
\newblock {EventEpi-A natural language processing framework for event-based surveillance}.
\newblock \emph{PLoS Computational Biology}, 16\penalty0 (11), 2020.
\newblock \doi{10.1371/journal.pcbi.1008277}.

\bibitem[Brown et~al.(2020)Brown, Mann, Ryder, Subbiah, Kaplan, Dhariwal, et~al.]{Brown2020}
T.~B. Brown, B.~Mann, N.~Ryder, M.~Subbiah, J.~Kaplan, P.~Dhariwal, et~al.
\newblock Language models are few-shot learners.
\newblock In \emph{Advances in Neural Information Processing Systems}, volume 2020-December, 2020.

\bibitem[Brownstein et~al.(2023)Brownstein, Rader, Astley, and Tian]{Brownstein20231597}
J.~S. Brownstein, B.~Rader, C.~M. Astley, and H.~Tian.
\newblock {Advances in Artificial Intelligence for Infectious-Disease Surveillance}.
\newblock \emph{New England Journal of Medicine}, 388\penalty0 (17):\penalty0 1597 – 1607, 2023.
\newblock \doi{10.1056/NEJMra2119215}.

\bibitem[Consoli et~al.(2019)Consoli, {Reforgiato Recupero}, and Petkovic]{DBLP:books/sp/CRP2019}
S.~Consoli, D.~{Reforgiato Recupero}, and M.~Petkovic, editors.
\newblock \emph{{Data Science for Healthcare - Methodologies and Applications}}.
\newblock Springer Nature, 2019.
\newblock \doi{10.1007/978-3-030-05249-2}.

\bibitem[Dong et~al.(2023)Dong, Li, Dai, Zheng, Wu, Chang, et~al.]{dong2023survey}
Q.~Dong, L.~Li, D.~Dai, C.~Zheng, Z.~Wu, B.~Chang, et~al.
\newblock A survey on in-context learning.
\newblock \emph{arXiv 2301.00234}, 2023.

\bibitem[Leuba et~al.(2020)Leuba, Yaesoubi, Antillon, Cohen, and Zimmer]{Leuba2020}
S.~I. Leuba, R.~Yaesoubi, M.~Antillon, T.~Cohen, and C.~Zimmer.
\newblock {Tracking and predicting U.S. influenza activity with a real-time surveillance network}.
\newblock \emph{PLoS Computational Biology}, 16\penalty0 (11), 2020.
\newblock \doi{10.1371/journal.pcbi.1008180}.

\bibitem[Madoff and Woodall(2005)]{madoff2005internet}
L.~C. Madoff and J.~P. Woodall.
\newblock {The internet and the global monitoring of emerging diseases: lessons from the first 10 years of ProMED-mail}.
\newblock \emph{Archives of Medical Research}, 36\penalty0 (6):\penalty0 724--730, 2005.

\bibitem[McDonald et~al.(2021)McDonald, Bien, Green, Hu, DeFries, Hyun, et~al.]{McDonald2021}
D.~J. McDonald, J.~Bien, A.~Green, A.~J. Hu, N.~DeFries, S.~Hyun, et~al.
\newblock {Can auxiliary indicators improve COVID-19 forecasting and hotspot prediction?}
\newblock \emph{Proceedings of the National Academy of Sciences of the United States of America}, 118\penalty0 (51), 2021.
\newblock \doi{10.1073/pnas.2111453118}.

\bibitem[Mukherjee et~al.(2023)Mukherjee, Mitra, Jawahar, Agarwal, Palangi, and Awadallah]{mukherjee2023orca}
S.~Mukherjee, A.~Mitra, G.~Jawahar, S.~Agarwal, H.~Palangi, and A.~Awadallah.
\newblock {Orca: Progressive Learning from Complex Explanation Traces of GPT-4}, 2023.

\bibitem[Sagi and Rokach(2018)]{Sagi2018}
O.~Sagi and L.~Rokach.
\newblock {Ensemble learning: A survey}.
\newblock \emph{Wiley Interdisciplinary Reviews: Data Mining and Knowledge Discovery}, 8\penalty0 (4), 2018.
\newblock \doi{10.1002/widm.1249}.

\bibitem[Salathé et~al.(2012)Salathé, Bengtsson, Bodnar, Brewer, Brownstein, Buckee, et~al.]{salathe2012}
M.~Salathé, L.~Bengtsson, T.~J. Bodnar, D.~D. Brewer, J.~S. Brownstein, C.~Buckee, et~al.
\newblock Digital epidemiology.
\newblock \emph{PLoS computational biology}, 8\penalty0 (7):\penalty0 e1002616, 2012.

\bibitem[Sutskever et~al.(2014)Sutskever, Vinyals, and Le]{sutskever2014sequence}
I.~Sutskever, O.~Vinyals, and Q.~V. Le.
\newblock Sequence to sequence learning with neural networks.
\newblock In \emph{Advances in neural information processing systems}, pages 3104--3112, 2014.

\bibitem[Touvron et~al.(2023)Touvron, Martin, Stone, Albert, Almahairi, Babaei, et~al.]{touvron2023llama}
H.~Touvron, L.~Martin, K.~Stone, P.~Albert, A.~Almahairi, Y.~Babaei, et~al.
\newblock Llama 2: Open foundation and fine-tuned chat models, 2023.

\bibitem[Vaswani et~al.(2017)Vaswani, Shazeer, Parmar, Uszkoreit, Jones, Gomez, Kaiser, and Polosukhin]{Vaswani20175999}
A.~Vaswani, N.~Shazeer, N.~Parmar, J.~Uszkoreit, L.~Jones, A.~Gomez, L.~Kaiser, and I.~Polosukhin.
\newblock Attention is all you need.
\newblock In \emph{Advances in Neural Information Processing Systems}, pages 5999--6009, 2017.

\bibitem[Vespignani(2011)]{vespignani2011}
A.~Vespignani.
\newblock Modelling dynamical processes in complex socio-technical systems.
\newblock \emph{Nature Physics}, 8\penalty0 (1):\penalty0 32--39, 2011.

\end{thebibliography}

\end{document}